\documentclass[12pt]{article}
\usepackage{bbm,amsmath,graphicx,setspace,amssymb}
\usepackage[font=small]{caption}
\begin{document}
\title{\bf Disperse two-phase flows, with applications to geophysical
  problems} 
\author{Luigi C. Berselli
  \\
  Dipartimento di Matematica
  \\
  Universit\`a di Pisa
  \\
  Pisa, ITALY, berselli@dma.unipi.it 
  \and
  Matteo Cerminara
  \\
  Scuola Normale Superiore di Pisa
  \\
  Istituto Nazionale di Vulcanologia e Geofisica
  \\
  Pisa, ITALY, matteo.cerminara@sns.it 
  \and 
  Traian Iliescu
  \\
  Department of Mathematics
  \\
  Virginia Tech
  \\
  Blacksburg, VA, iliescu@vt.edu}
\date{}
\maketitle
\begin{abstract}
  In this paper we study the motion of a fluid with several dispersed
  particles whose concentration is very small (smaller than
  $10^{-3}$), with possible applications to problems coming from
  geophysics, meteorology, and oceanography. We consider a very dilute
  suspension of heavy particles in a quasi-incompressible fluid (low
  Mach number).  In our case the Stokes number is small and --as
  pointed out in the theory of multiphase turbulence-- we can use an
  Eulerian model instead of a Lagrangian one.  The assumption of low
  concentration allows us to disregard particle--particle
  interactions, but we take into account the effect of particles on
  the fluid (two-way coupling). In this way we can study the physical
  effect of particles' inertia (and not only passive tracers), with a
  model similar to the Boussinesq equations.

  The resulting model is used in both direct numerical simulations and
  large eddy simulations of a dam-break (lock-exchange) problem, which
  is a well-known academic test case.
\end{abstract}
%
%\doublespacing
\section{Introduction}
One of the characteristic features of geophysical flows (see for
instance~\cite{CRB2011}) is stratification (the other one is
rotation).  In this manuscript, we study some problems related to
suspensions of heavy particles in incompressible \mbox{-or} slightly
compressible- fluids. Our aim is a better understanding of mixing
phenomena between the two phases, the fluid and solid one. We
especially study this problem because (turbulent) mixing with
stratification plays a fundamental role in the dynamics of both
oceanic and atmospheric flows. In this study, we perform the analysis
of some models related to the transport of heavy dilute particles,
with special emphasis on their mixing.  Observe that mixing is very
relevant near the surface and the bottom of the ocean, near
topographic features, near polar and marginal seas, as well as near
the equatorial zones~\cite{KC2000}.  Especially in coastal waters,
precise analysis of transport and dispersion is needed to study
biological species, coastal discharges, and also transport of
contaminants. The other main motivation of our study is a better
understanding of transport of particles (e.g. dust and pollution) in
the air. This happens -for instance- in volcanic eruptions or more
generally by natural and/or human generation of jets/plumes of
particles in the atmosphere.

Following~\cite{BE2010}, in the physical regimes we will consider, it
is appropriate to use the Eulerian approach, that is the solid-phase
(the particles) will be modeled as a continuum.  This choice is
motivated by the presence of a huge number of particles and because we
are analyzing the so called ``fine particle'' regime (that is the
Stokes number is much smaller than one). In this regime, a Lagrangian
approach could be computationally expensive, and the Eulerian approach
may offer more computationally efficient alternatives. We will explain
the precise assumptions that make this \textit{ansatz} physically
representative and we will also study numerically the resulting
models, with and without large scales further approximation. In
particular, we will model the particles as dust, investigating a model
related to \textit{dusty gases}, and which belongs to the hierarchy of
reduced multiphase models, as reviewed by Balachandar~\cite{BE2010}.
These models represent a good approximation when the number of fine
particles to be traced is very large and a direct numerical simulation
(DNS) of the fluid with a Lagrangian tracer for each particle would be
too expensive. As well explained in~\cite{BE2010}, the point-like
Eulerian approach for multiphase fluid-particle dynamics becomes even
more efficient in the case of large eddy simulations (LES), because
the physical diameter of the particles has to be compared with the
large eddy length-scale and not with the smaller Kolmogorov one. We
will use the dusty gas model in a physical configuration that is very
close to that modeled by the Boussinesq system, and this explains why
we compare our numerical results with those reported
in~\cite{OIFSD2007,BFIO2011}. Observe that the dusty gas model reduces
to the Boussinesq system with a large Prandtl number if: a) the fluid
velocity is divergence-free; and b) the relative ratio of solid and
fluid bulk densities is very small (see Sec.~\ref{sec:models} and
Eq.~\eqref{eq:Boussinesq}).

The approach we will use for multiphase fluids is well-described in
Marble~\cite{Mar1970}. More precisely, when the \textit{Stokes time}
--which is the characteristic time of relaxation of the particle
velocity with respect to the surrounding fluid-- is small enough and
the number of particles is very large, it could be reasonable to use
the Eulerian approach (instead of the Lagrangian). In Eulerian models
both the carrier and the dispersed phase are treated as
interpenetrating fluid media, and consequently both the particulate
solid-phase and fluid-phase properties are expressed by a continuous
field representation. Originally we started studying these models in
order to simulate ash plumes coming from volcanic eruptions,
see~\cite{EOCENS2007,CBEOS2013,Cer2014,VHC2014}, but here we will show
that the same approach could be also used to study some problems
coming from other geophysical situations, at least for certain ranges
of physical parameters.

Our model is evaluated in a two dimensional \textit{dam-break
  problem}, also known as the \textit{lock-exchange problem}. This
problem, despite being concerned with a) a simple domain; b) nice
initial and boundary conditions; and c) smooth gravity external
forcing, contains shear-driven mixing, internal waves, interactions
with boundaries, and convective motions.  The dam-break problem setup
has long served as a paradigm configuration for studying the
space-time evolution of gravity currents
(cf.~\cite{Dor1998,Fer2000,RL2000,SD1994}). Consequently, we set up a
canonical benchmark problem, for which an extensive literature is
available: The vertical barrier separating fluid and fluid with
particles is abruptly removed, and counter-propagating gravity
currents initiate mixing. The time evolution can be quite complex,
showing shear-driven mixing, internal waves interacting with the
velocity, and gravitationally-unstable transients.  This benchmark
problem has been investigated experimentally and numerically for
instance in~\cite{BS1978,HLD1996,HHPS1996,HPHS1993}. Both the
impressive amount of data and the physical relevance of the problem
make it an appropriate benchmark and a natural first step in the
thorough assessment of any approximate model to study
stratification. The results we obtain validate the proposed model as
appropriate to simulate dilute suspensions of ash in the air. In
addition, we found that new peculiar phenomena appear, which are
generated by compressibility. Even if the behavior of the simulations
is qualitatively very close to that of the incompressible case, the
(even very slightly) compressible character of the fluid produces a
more complex behavior, especially in the first part of the
simulations. To better investigate the efficiency and limitations of
the numerical solver, the numerical tests will be performed by using
both DNS and LES. Complete discussion of the numerical results will be
given in Section~\ref{sec:numerical_results}.

\medskip

\textbf{Plan of the paper:} In Section~\ref{sec:models} we present the
reduced multiphase model we will consider, with particular attention
to the correct evaluation of physical parameters that make the
approximation effective. In Section~\ref{sec:numerical_results} we
present the setting of the numerical experiments we
performed. Particular emphasis is posed on the initial conditions and
on the interpretation and comparison of the results with those
available in the literature.
\section{On multiphase Eulerian models}
\label{sec:models}
In order to study multiphase flows and especially (even compressible)
flows with particles, some approximate and reduced models have been
proposed in the literature. In the case of dilute suspensions, a
complete hierarchy of approximate models is available
(see~\cite{BE2010}) on the basis of two critical parameters
determining the level of interaction between the liquid and solid
phase: The fractional volume occupied by the dispersed-phase and the
mass loading, (that is the ratio of mass of the dispersed to carrier
phase).  When they are both small, the dominant effect on the dynamics
of the dispersed-phase is that of the turbulent carrier flow
(\textit{one-way coupled}). When the mass of the dispersed-phase is
comparable with that of the carrier-phase, the back-influence of the
dispersed-phase on the carrier-phase dynamics becomes relevant
(\textit{two-way coupled}). When the fractional volume occupied by the
dispersed-phase increases, interactions between particles become more
important, requiring a \textit{four-way coupling}. In the extreme
limit of very large concentration, we encounter the granular flow
regime.

Here, we consider rather heavy particles such that
$\widehat{\rho}_{s}>>\widehat{\rho}_{f}$ (air), or
$\widehat{\rho}_{s}\sim\widehat{\rho}_{f}$ (liquid), where in the
sequel the subscript ``$_s$'' stands for solid, while ``$_f$'' stands
for fluid.  Here a hat $\,\widehat{.}\,$ denotes material densities
(as opposed to bulk densities): In particular, we suppose
$\widehat{\rho}_{s}\sim 400-3000 \text{kg}/\text{m}^{3}$. A rather
small particle/volume concentration must be assumed (to have dilute
suspensions), that is
\begin{equation*}
  \epsilon_s := \frac{V_{s}}{V} < 10^{-3},
\end{equation*}
where $V_{s}$ is the volume occupied by the particles over the total
volume $V$. When $\epsilon_{s}$ is smaller than $10^{-3}$,
particle-particle collisions and interactions can be neglected and the
particle-phase can be considered as a pressure-less and non-viscous
continuum. In this situation the particles move approximately with the
same velocity of the surrounding fluid, and the theory has been
developed by Carrier~\cite{Car1958} (see a review in
Marble~\cite{Mar1970}).  With these assumptions the bulk densities
$\rho_{f}:=(1-\epsilon_{s})\widehat{\rho}_{f}$ and
$\rho_{s}:=\epsilon_{s}\widehat{\rho}_{s}$ are of the same order of
magnitude, about $1 \text{kg}/\text{m}^{3}$ in the case dust-in-air
(two-way coupling). In the case of water with particles the ratio
$\rho_s/\rho_f$ is of the order of $10^{-3}$, hence particles behave
very similarly to passive tracers (almost one-way coupling).

Another assumption required by Marble's analysis is that particles can
be considered \textit{point-like}, if their typical diameter $d_s$ is
smaller than the smallest scale of the problem under analysis, that is
the Kolmogorov length $\eta$ (DNS), or the smallest resolved LES
length-scale $\xi$ (LES).

To describe the gas/fluid-particle drag, we observe that it depends in
a strong nonlinear way on the local flow variables and especially on
the relative Reynolds number:
\begin{equation*}
  \text{Re}_s= \frac{\widehat{\rho}_f |\,{u}_s - \,{u}_f| d_s}{\mu},
\end{equation*}
where $\mu$ is the gas dynamic viscosity coefficient and $\,{u}_{f}$
and $u_s$ are the fluid and solid phase velocity field, respectively.
On the other hand, for a point-like single particle and in the
hypothesis of small velocities difference ($\text{Re}_s<1$), the drag
force (per volume unit) acting on a single particle depends just
linearly on the difference of velocities:
\begin{equation*}
  {f}_d = \frac{\rho_s}{\tau_{s}}(\,{u}_s - \,{u}_f)\,,\qquad
  \text{with}\qquad \tau_{s} :=\frac{({2
      \widehat{\rho}_{s}/\widehat{\rho}_{f}+1})\,d_s^2}{36\nu}, 
\end{equation*}
where $\tau_{s}$ is the \textit{particle relaxation time or Stokes
  time}, which is the time needed to a particle to equilibrate to a
change of fluid velocity~\cite{BE2010}, and
$\nu:={\mu}/{\widehat{\rho}_f}$ is the fluid kinematic viscosity. In
particular, in the case of water with particles we have
$\widehat{\rho}_{s}/\widehat{\rho}_{f}\sim1$, while in the case of a
gas $\widehat{\rho}_{s}/\widehat{\rho}_{f}>>1$ and hence
\begin{equation}
\label{eq:tau_w}
    \tau_{s} \ \sim\ 
  \left\{\begin{aligned}
      &\quad \frac{d_s^2}{12\nu} \qquad \text{(water),}   
      \\
      &\quad\frac{\widehat{\rho}_s\,d_s^2}{18\mu} \qquad \text{(air).} 
 \end{aligned}\right.
\end{equation}
In order to measure the lack of equilibrium between the two phases, we
have to compare $\tau_{s}$ with the smallest time of the dynamics. In
the turbulent regime, the smallest time is the Kolmogorov smallest
eddy's turnover time $\tau=\tau_{\eta}=\eta^2/\nu$ (DNS)
(cf. Frisch~\cite{Fri1995}) or analogously
$\tau=\tau_\xi=\tau_{\eta}\,(\xi/\eta)^\frac{2}{3}$ (LES).  It is
possible to characterize this situation by using as non-dimensional
parameter --the Stokes number-- which is defined by comparing the Stokes
time with the fastest time-scale of the problem under analysis
$\text{St} :={\tau_{s}/}{\tau}$.  If $\textup{St} < 10^{-3}$ (the
``fine particle regime''), we say that we have \textit{kinematic
  equilibrium} between the two phases and so we can use in a
consistent way the dusty gas model. In order to have also
\textit{thermal equilibrium} between the two phases, one has to assume
that the \textit{thermal relaxation time} (cf.~\cite{Mar1970}) is
small, that is:
\begin{equation*}
  \tau_T := \frac{\widehat{\rho}_s C_s}{k_s} \frac{d_s^2}{4}<<1. 
\end{equation*}
Comparing the kinetic and thermal relaxation times, we get the Stokes
thermal time
\begin{equation}
  \label{eq:Stokes}  
  \text{St}_{T}:=\frac{\tau_T}{\tau} =
  \frac{\tau_T}{\tau_s}\frac{\tau_s}{\tau}=\frac{3}{2} \frac{C_s 
    \mu}{k_s} \text{St}= \frac{3}{2} \,{Pr}_s \,\text{St}\,,
\end{equation}
i.e., the particle Prandtl number, where $C_s$ is the solid-phase
specific heat-capacity at constant volume and $k_s$ is its thermal
conductivity. To ensure that the dusty gas model is physically
reasonable, both kinematic and thermal equilibrium must hold, that is,
both Stokes numbers should be less than $10^{-3}$.  This implies that
we have a single velocity $u=u_f=u_s$ for both phases and also a
single temperature field $T=T_f=T_s$.

To check that our assumptions are fulfilled, we first show that if the
Stokes number is small, then also the thermal Stokes number remains
small. Indeed, using the typical value of the dynamic viscosity
$\mu=10^{-3}\,\text{Pa}\cdot\text{s}$ (water) or
$\mu=10^{-5}\,\text{Pa}\cdot\text{s}$ (air), specific heat capacity
$C_s=10^3\,\mathrm{J \cdot kg^{-1} \cdot K^{-1}}$ and thermal
conductivity $k_s\sim1\,\mathrm{W}\cdot \text{m}^{-1}\cdot
\text{K}^{-1}$, we can evaluate the particle Prandtl number in both
cases:
\begin{equation*}
  \text{Pr}_s =\frac{\mu C_s}{k_s}\sim 
  \left\{\begin{aligned}
      \frac{10^3 * 10^{-3}}{1} \sim 1 \qquad \text{(water),}
      \\
      \frac{10^3 * 10^{-5}}{1} \sim 10^{-2} \qquad \text{(air).}
 \end{aligned}\right.
\end{equation*}
Hence formula~\eqref{eq:Stokes} shows that
$\text{St}_T\lesssim\text{St}$.

Summarizing, we used the following assumptions:
\begin{enumerate}
\item[a)] Continuum assumption for both the gaseous and solid phase;
  \vspace{-.5cm}
\item[b)] The solid-phase is dispersed ($\epsilon_s < 10^{-3}$), thus
  it is pressure-less and non-interacting; \vspace{-.5cm}
\item[c)] The relative Reynolds number between the solid and gaseous
  phases is smaller than one so that it is appropriate to use the
  Stokes law for drag; 
  \vspace{-.5cm}
\item[d)] The Stokes number is smaller than one so that the Eulerian
  approach is appropriate; 
  \vspace{-.5cm}
\item[e)] All the phases, either solid or gaseous, have the same
  velocity and temperature fields $u(x,z,t),\,T(x,z,t)$ (local thermal and
  kinematic equilibrium). We showed that this assumption is accurate
  if the Stokes number is much smaller than one.
\end{enumerate}
In this regime, the equations for the balance of mass, momentum, and
energy are:
\begin{equation}
  \label{eq:equilibriumEulerian2}
  \left\{
    \begin{aligned}
      &\partial_t\rho + \nabla\cdot(\rho \,{u}) = 0,
      \\
      &\partial_t\rho_s + \nabla\cdot(\rho_s \,{u}) = 0,
      \\
      &\partial_t (\rho \,{u}) + \nabla\cdot (\rho \,{u}\otimes \,{u}
      + p\,\mathbbm{I} - \mathbbm{T}) =
      % \\
      % \qquad\qquad\qquad\qquad\quad 
       \rho  \,{g},
      \\
      & \partial_t (\rho\, e) + \nabla\cdot\left(\rho\,u\, e\right) + p\,\nabla\cdot\,{u} = 
      % \\
      % \qquad\; 
      \mathbbm{T}:\nabla \,{u}-\nabla\cdot q\,,% \tau_{S}\left[\rho_s| {a}|^2 +
      % \nabla\cdot\left(\rho_s\frac{C_as}{C_v}e_f\vec{a}\right)\right]\,,
    \end{aligned}
  \right.
\end{equation}
where $\rho := \rho_f + \rho_s$ is the mixture density,
$e:=\frac{C_v \rho_f +{C_s}\rho_s}{\rho}$ is the internal mixture
energy, and $g$ is the gravity acceleration pointing in the downward vertical direction.  The
stress-tensor is
\begin{equation*}
 \mathbbm{T} :=
  2\mu(T)\,\left[\frac{\nabla{u}+\nabla{u}^{T}}{2}-
  \frac{1}{D}(\nabla\cdot{u})\,\mathbbm{I}\right],
\end{equation*}
with $\mu(T)$ the dynamic viscosity, possibly depending on the
temperature $T$, and $D$ the spatial dimension of the problem. The
Fourier law for the heat transfer assumes $\vec{q} = -k\,\nabla T$,
where $k$ is the fluid thermal conductivity. We denote by $C_v$ and
$C_s$ the fluid and solid phase specific heat-capacity at constant
volume, respectively.  System~\eqref{eq:equilibriumEulerian2} is
completed by using the constitutive law $p=p(\rho,\rho_s,T)$. In the
case of air and particles (the one for which we will present the
simulations) $p=\rho_f R\,T$, where $R$ is the air gas constant.
\\
\textbf{Remark 1.}  The correct law would be $p=\frac{\rho_f
  R\,T}{1-\epsilon_s}$, but in our dilute setting $\epsilon_s$ is very
small, which justifies the approximation $p=\rho_f R\,T$. A different
constitutive law must be used in the presence of water or other
fluids.
\\
\textbf{Remark 2.}  Note that the constant particle pressure $\nabla
p_s=0$ is justified by the lack of particle-particle forces.  Note
that in the case of uniform particle distribution ($\rho_s/\rho_f=
C$), the equations~\eqref{eq:equilibriumEulerian2} reduce to the
compressible Navier-Stokes equations, with density multiplied by a
factor $C$.  Some numerical experiments (with $\rho_s/\rho_f\not= C$)
were performed in~\cite{SKOH2005}, where the dusty gas model was
applied to volcanic eruptions, i.e. a flow with vanishing initial
solid density $\rho_s$ and particles injected into the atmosphere from
the volcanic vent.

Denoting by s $y_s={\rho_s}/{\rho}$ the solid-phase mass-fraction, we
can rewrite the system~\eqref{eq:equilibriumEulerian2} with just one
flow variable ($\rho\,u$) as follows:
\begin{equation}
  \label{eq:equilibriumEulerian2bis}
  \left\{
    \begin{aligned}
      &\partial_t\rho + \nabla\cdot(\rho \,{u}) = 0,
      \\
      &\partial_t(\rho\, y_s) + \nabla\cdot(\rho\,{u}\,y_s) = 0,
      \\
      &\partial_t (\rho \,{u}) + \nabla\cdot (\rho \,{u}\otimes
      \,{u} + p\,\mathbbm{I} - \mathbbm{T}) =  
      % \\
      % \qquad\qquad\qquad\qquad\quad 
       \rho  \,{g},
      \\
      & \partial_t (\rho\, e) + \nabla\cdot\left(\rho\,{u}\, e\right) + p\,\nabla\cdot\,{u} = 
      % \\
      % \qquad\; 
      \mathbbm{T}:\nabla \,{u}-\nabla\cdot q\,.% \tau_{S}\left[\rho_s| {a}|^2 +
      % \nabla\cdot\left(\rho_s\frac{C_as}{C_v}e_f\vec{a}\right)\right]\,,
    \end{aligned}
  \right.
\end{equation}
In the following, we will also assume to have an iso-entropic flow
with a perfect gas (which is a reasonable approximation for the air,
see for example~\cite{M1956}).  We can thus substitute the energy
equation~(\ref{eq:equilibriumEulerian2bis}-d) by the constitutive
law 
\begin{equation*}
p(x(t),z(t),t)=p_0(x(0))\left(\frac{\rho(x(t),z(t),t)}{\rho(x(0),z(0),0)}\right)^{\gamma(x(t),z(t),
t)}, 
\end{equation*}
where $\gamma(x(t),z(t),t)=\frac{1-y_s(x(t),z(t),t)
  R}{(1-y_s(x(t),z(t),t))C_v+y_s(x(t),z(t),t) C_s}$ and $(x(t),z(t))$
is the streamline starting at $(x(0),z(0))$ for $t=0$ (we have not
been able to find this expression in the literature; for its full
derivation see~\cite{Cer2014}). In particular, a simple calculation
shows that $\gamma(x(t),z(t),t)=\gamma(x(0),z(0),0)\sim
\gamma$. Moreover, since
$T(x(0),z(0),0)/\rho(x(0),z(0),0)^{\gamma(x(0),z(0),0)}=a(x(0),z(0),0)\sim
a$ (where $a(x(0),z(0),0)\sim a$ and $\gamma(x(0),z(0),0)\sim \gamma$
are motivated by the small density variations compared with a constant
temperature), we can consequently study the following system (with
$p=a\, \rho^\gamma$; and $a,\,\gamma$ are constants determined from
the initial conditions):
\begin{equation}
  \label{eq:equilibriumEulerian-iso}
  \left\{
    \begin{aligned}
      &\partial_t\rho + \nabla\cdot(\rho \,{u}) = 0,
      \\
      &\partial_t(\rho\, y_s) + \nabla\cdot(\rho\, y_s \,{u}) = 0,
      \\
      &\partial_t (\rho \,{u}) + \nabla\cdot (\rho \,{u}\otimes \,{u}
      + %a \,\rho^\gamma
      p\,\mathbbm{I} - \mathbbm{T}) =
      % \\
      % \qquad\qquad\qquad\qquad\quad 
       \rho\,  {g}.
    \end{aligned}
  \right.
\end{equation}
Here the iso-entropic assumption is justified. Indeed, since the
Reynolds number is typically much greater than 1, and the Prandtl
number is of the order of $10$, the two dissipation terms
$\mathbbm{T}:\nabla \,{u}$ and $\nabla\cdot q$ (corresponding to the
conduction of heat and its dissipation by mechanical energy) can be
neglected. Moreover, since $C_v\sim C_s$ and the temperature
fluctuations are small, we can disregard the heat transfer from solid
to fluid phase.

Observe that if $\rho_f=\textup{constant}$, $T=\textup{constant}$, and
if we use the Boussinesq approximation, we get
from~\eqref{eq:equilibriumEulerian2bis} the following system:
\begin{equation}
  \label{eq:Boussinesq}
  \left\{
    \begin{aligned}
      &\nabla\cdot {u} = 0,
      \\
      &\partial_t\rho_s + (u\cdot \nabla)\, \rho_s = 0,
      \\
      &\partial_t \,{u} + \nabla\cdot (\,{u}\otimes \,{u} +
      p\,\mathbbm{I} - \mathbbm{T}) = \rho_s {g}\,,
    \end{aligned}
  \right.
\end{equation}
which is exactly the Boussinesq equations, except that there is no
diffusion for the density perturbation (i.e., infinite Prandtl
number). Thus, numerical results
concerning~\eqref{eq:equilibriumEulerian-iso} are comparable with
results from the classical Boussinesq equations,
see~\cite{OIFSD2007,BFIO2011}.
\section{Numerical results}
\label{sec:numerical_results}
To validate the Eulerian model for multiphase
flows~\eqref{eq:equilibriumEulerian-iso}, we use it to perform both
DNS and LES of a dam-break (lock-exchange) problem.
\subsection{Model configuration}
Since we want to compare our results with accurate results available
in the literature, we use a setting which is very close to that
in~\cite{OIFSD2007}, in terms of both equations and initial
conditions. In particular, we consider a two dimensional rectangular
domain $-L/2\leq x\leq L/2$ and $0\leq z \leq H$ with an aspect ratio
large enough ($L/H=5$) in order to obtain high shear across the
interface, and to create Kelvin-Helmholtz (KH) instability. We use
this setting because in a domain with large aspect ratio, the density
interface has more space to tilt and stretch.

For this test case, the typical velocity magnitude is (for further
details see e.g.~\cite{OIFSD2007}) $U_0=\sqrt{g\rho_s h (H - h)/\rho_0
  H}$, where $H$ is the layer thickness and $h$ the volumetric
fraction of denser material times $H$. From now on, with a slight
abuse of notation, we denote by $g$ the modulus of the gravity
acceleration. In our simulation we set $h=H/2$, from which we get
\begin{equation*}
  U_0 = \frac{1}{2}\,\sqrt{\frac{g \rho_s H}{\rho_0}}.
\end{equation*}
We use the characteristic length-scale $\ell$ to non-dimensionalize
all the equations in~\eqref{eq:equilibriumEulerian-iso}. In order to
have $\tau = \ell/U_0$ when $H=2\ell$, we need to set $\rho_0/g =
\rho_s/2$. Moreover, we choose a dimensional system where the initial
solid bulk density is $\rho_{s,0} = 1$, which yields $g =
2\rho_0$. The Froude number is $2^{-1/2}$ for all the simulations, so
we are free to choose a $\rho_0$ such that $\rho_0 \gg \rho_s$. We set
$\rho_0 =100$.  In these non-dimensional units, the Reynolds number is
$Re = (\rho+\rho_s)U_0\ell/\mu = (\rho_0 + 1)/\mu$, we set the dynamic
viscosity $\mu=0.02348837$ such that the maximum Reynolds number we
consider is
\begin{equation*}
  Re =4300.
\end{equation*}
One of the inherent time-scales in the system is the
(Brunt-V\"ais\"al\"a) buoyancy period
\begin{equation*}
  T_b = 2\pi\sqrt{\frac{\rho_0 H}{g \rho_s}} = 2\pi,
\end{equation*}
which is the natural time related to gravity waves. In order to have a
quasi-incompressible flow, we set $\textup{Ma}=U_0/c=0.01$. Using our
non-dimensional variables, the perfect gas relationship is $p_0=\rho_0
R$ and the speed of sound is $c=\gamma R$. We want $c=100$ and
$\gamma=1.4$, so we set $R=7142.857143$ and
$p_0=7.142857143*10^5$. Experiments are performed at different
resolutions (from about $10^4$, up to about $10^6$ grid cells), see
the next section for details.

The initial condition is a state of rest, in which the fluid with
particles on the left is separated from the fluid (without particles)
on the right by a sharp transition layer. Since the tilting of the
density interface puts the system gradually into motion, the system
can be started from a state of rest. Due to the (slight)
compressibility of the fluid some peculiar phenomena occur close to
the initial time. These effects are not present in the incompressible
case, cf.~the discussion below.

We consider the isolated problem, so that the iso-entropic
approximation is valid and consequently we supplement
system~\eqref{eq:equilibriumEulerian-iso} with the following boundary
conditions: The boundary condition for the density perturbation $y_s$
is no-flux, while free-slip for the velocity:
 \begin{equation*}
   \left\{   
     \begin{aligned}
       & u\cdot n=0,
       \\
       &n\cdot (\mathbbm{T}-p\,\mathbbm{I})\cdot \tau=0,
     \end{aligned}
   \right.
   \qquad \text{and}\qquad n\cdot\nabla y_s=0,
 \end{equation*}
 where $n$ is the unit outward normal vector, while $\tau$ is a
 tangential unit vector on $\partial \Omega$.  In the two dimensional
 setting we use for the numerical simulation (the two dimensional
 rectangular domain $\Omega=]-L/2,L/2[\times]0,H[$) the boundary
 conditions become:
 \begin{equation*}
   \left\{
     \begin{aligned}
       & \frac{\partial u_1}{\partial z}=0, \quad u_2=0,\quad
       \frac{\partial y_s}{\partial z}=0,\quad \text{ at }z=0,H,\
       -\frac{L}{2}<x<\frac{L}{2},
       \\
       & \frac{\partial u_2}{\partial x}=0, \quad u_1=0,\quad
       \frac{\partial y_s}{\partial x}=0, \quad \text{ at }x=\pm
       \frac{L}{2},\ 0<z<H.
     \end{aligned}
   \right. 
\end{equation*}
\subsection{On the initial conditions}
We considered as initial datum the classical situation used in the
dam-break problem, with all particles confined in the left half of the
physical domain (with uniform distribution), while a uniform fluid
fills the whole domain. Moreover, we have an initial uniform
temperature $T(x,z,0)=T_{0}$ and pressure distribution
$p(x,z,0)=p_{0}$. Suddenly the wall dividing the two phases is removed
and we observe the evolution.

Even if our numerical code is compressible, we started with this
setting, widely used to study incompressible cases, since we are in
the physical regime of quasi-incompressibility. The compressibility is
mostly measured by the Mach number. For air we have a typical velocity
$U_0\sim 4m/s$, hence the Mach number of air in this condition is
around 0.01, as we choose for our simulations. On the other hand, for
water we would obtain $U_0\sim 0.04m/s$ and $\textup{Ma}\sim 2.5\
10^{-5}$. Nevertheless, as we will see especially in
Fig.~\ref{fig:BGE}, even this very small perturbation creates a new
instability and new phenomena for times very close to $t=0$. In
particular, new effects appear for $0 < t < T_b$.  These effects seem
limited to the beginning of the evolution. The characteristic time of
the stratification (for a DNS) is defined as (see~\cite[\S~11]{CRB2011})
\begin{equation*}
  T_a=2\pi\sqrt{\frac{\rho H}{g \Delta\rho_f}},
\end{equation*}
where $\Delta \rho_f$ is the density difference between the ground
level and the height $H$ of the upper boundary wall. In particular, we
know that for the gaseous-phase, the stable solution is the barotropic
stratification, due to the gravity acceleration:
\begin{equation*}
T(z)=T_{0}-\frac{g\,z}{\gamma C_{v}},\quad
\rho(z)=\rho_{0}\left(\frac{T(z)}{T_{0}}\right)^{\frac{1}{\gamma-1}},
\quad 
p(z)=\rho_{0}\left(\frac{T(z)}{T_{0}}\right)^{\frac{\gamma}{\gamma-1}},
\end{equation*}
and in the case of perfect gases we recover the fact
that the typical stratification height for the atmosphere ($R\sim287$) is 
\begin{equation*}
  z_{gas}=\frac{1}{  \eta_{gas}}=\frac{\gamma\,R\,T}{g}\sim 10^4m,
\end{equation*}
while for water in the iso-thermal case we would obtain
\begin{equation*}
  z_{water}=\frac{1}{  \eta_{water}}=\frac{1}{\alpha \rho_0 g}\sim 10^5m.
\end{equation*}
Since $\eta$ is small in both cases, we can use the following
approximation:
\begin{equation*}
  \rho(z)\sim\rho_{0}\left(1-\frac{g\,z}{\gamma\,R\,T_{0}}\right)
  :=\rho_{0}\left(1-\eta \,z\right)\,.
\end{equation*}
For a domain with volume $V$ and mass $m$, in the incompressible case 
the stable stationary configuration is with vanishing velocity and
$\rho_{homog.}=\frac{m}{V}$.  On the contrary, in our slightly
compressible case, the stable stationary configuration is:
\begin{equation*}
  \frac{\rho(z)}{\rho_{homog.}}=\frac{1-\eta\, z}{1-\eta\,\frac{ {H}}{2}}.
\end{equation*}
The length $\eta$ has to be compared with the height of the domain
$H$, in order to evaluate the importance of stratification. For
instance, if we use realistic values of density, pressure, and gravity
acceleration for air (to come back to dimensional variables) we get
that the height of the domain is $H_{air}\sim 600m$, while for water
we get $H_{water}=0.6m$. In the case of air we obtain density
variations due to gravity which are of the order of $5\%$, while for
water they should be of the order of $0.0003\%$.  This explains that
in the case of water, the dominant variations of density, which are of
the order of $1\%$, are those imposed by the initial configuration of
particles. On the other hand, in the case of particles in air, the one
we are mostly interested to, the two phenomena create fluctuations
which are comparable in magnitude, and this can be seen in
Fig.~\ref{fig:BGE_irreversible}. In particular, in
Fig.~\ref{fig:BGE_irreversible}, one can see that the fluctuations
created by the non-stratified initial condition affect the behavior of
the background potential energy defined below. In the case of air, we
have that $T_a<T_b$, and thus the effects of these instabilities (due
to the initial heterogeneity) will be observed before the mixing
effects, which are dominant in the rest of the evolution. On the other
hand, this effect can not be seen by analyzing just the mixed
fraction, see Fig.~\ref{fig:mass-fraction} and the discussion below.

We will also compare the results obtained from DNS with those obtained
by different LES models, as discussed later on. The accuracy of the
LES models is evaluated through \textit{a posteriori} testing. The
main measure used is the background/reference potential energy (RPE),
which represents an appropriate measure for mixing in an enclosed
system~\cite{WLRD1995}.  RPE is the minimum potential energy that can
be obtained through an adiabatic redistribution of the masses. To
compute RPE, we use directly the approach in~\cite{WLRD1995}, since
the problem is two-dimensional and computations do not require too
much time
\begin{equation*}
  RPE(t) := g  \int_\Omega \rho_s(x,z,t) \,z_r(x,z,t)\,dxdz, 
\end{equation*}
where $z_r(\rho')$ is the height of fluid of density $\rho'$ in the
minimum potential energy state.  
To evaluate $z_r(\rho')$, we use the following formula:
\begin{equation*}
  z_r(x,z,t)=\frac{1}{L}\int_\Omega
  \mathcal{H}(\rho_s(x',z',t)-\rho_s(x,z,t))\, dx'dz',
\end{equation*}
where $\mathcal{H}$ is the Heaviside function. It is convenient to use
the non-dimensional background potential energy
\begin{equation}
  RPE^*(t):= \frac{RPE(t)-RPE(0)}{RPE(0)}\,, 
  \label{rpenondimensional}
\end{equation}
which shows the relative increase of the RPE with respect to the
initial state by mixing. Further discussion of the energetics of the
dam-break problem can be found
in~\cite{CRB2011,OIF2009b,OIF2009,OIFSD2007}.

With these considerations we are now able to compute the maximum
particle diameter fulfilling our hypothesis
($\textup{St}<10^{-3}$). First, we must evaluate the smallest
time-scale of the dynamics.  As described in
Tab.~\ref{tab:resolutions}, we used three different resolutions. The
ultra-res resolution can be considered as a DNS, so the smallest
time-scale of the simulation is the Kolmogorov time $\tau_\eta =
\textup{Re}^{-\frac{1}{2}} = 1.525*10^{-2}$, while the smallest
length-scale is $\eta = \textup{Re}^{-\frac{3}{4}}=1.883*10^{-3}$. The
other two resolutions have been used for LES: We have $\xi =
8.696*10^{-3}$ and $\xi = 4.348*10^{-2}$ for the mid-res and low-res
resolutions, respectively. By using the relationship $\tau_\xi =
\tau_\eta (\xi/\eta)^{\frac{2}{3}}$, we found $\tau_\xi =
4.229*10^{-2}$ and $\tau_\xi = 1.237*10^{-1}$, respectively. In
Tab.~\ref{tab:diameter} we report the dimensional maximum particle
diameter for which the dusty gas hypothesis is fulfilled
(cf. Eqs.~\eqref{eq:tau_w}) at various resolutions.
\begin{table}[h!]
\centering
\begin{tabular}{|c|c|c|c|}
  \hline
  & ultra-res & mid-res & low-res
  \\
  water & 6.3 \textup{$\mu$m} & 10 \textup{$\mu$m} & 18
  \textup{$\mu$m} 
  \\
  gas & 82 \textup{$\mu$m} & 140 \textup{$\mu$m} & 240 \textup{$\mu$m}
  \\
  \hline
\end{tabular}
\caption{The dimensional maximum particle diameter fulfilling the
  dusty gas hypothesis.} 
\label{tab:diameter}
\end{table}
\subsection{Numerical methods and results}
We tested our numerical code on a well documented test case. At the
initial time the particles occupy only one side of the computational
domain. Then --abruptly-- the wall dividing the fluid with particles
from the fluid without particles is removed and the two fluids start
mixing under the effect of gravity. The situation is complex even in
the two dimensional case.  Results of numerical simulations with the
DNS and also LES models are presented in this section. All simulations
are obtained by using OpenFOAM$^\textup{\textregistered}$, which is an
Open Source computational fluid dynamics code used worldwide. The
numerical algorithm we used is PISO (Pressure Implicit with Splitting
of Operators~\cite{FP1999,Iss1986}), which allows the user to choose
the numerical scheme and order for both the time and space
discretization. In particular, we choose a second order unbounded and
conservative scheme for the Laplacian terms; a central second order
scheme for interpolation from cell center to cell faces; a second
order scheme for the gradient terms; and a bounded second central
scheme for the divergence term~\cite{Jas1996}. On the other hand, we
choose a second order bounded and implicit time scheme
(Crank-Nicolson), with an adaptive time stepping based on the maximum
initial residual of the previous time step~\cite{KGGS}, and on a
threshold that depends on the Courant number ($\textup{C} < 0.2$).

The linear system is solved by using the PbiCG solver (Preconditioned
bi-Conjugate Gradient solver for asymmetric matrices) and the PCG
(Preconditioned Conjugate Gradient solver for symmetric matrices),
respectively, preconditioned by a Diagonal Incomplete Lower Upper
decomposition (DILU) and a Diagonal Incomplete Cholesky (DIC)
decomposition. The tolerance has been set to $0.01$ for the initial
residual and to $10^{-15}$ for the final one.

\begin{figure}
\centering

\includegraphics[width=\columnwidth]{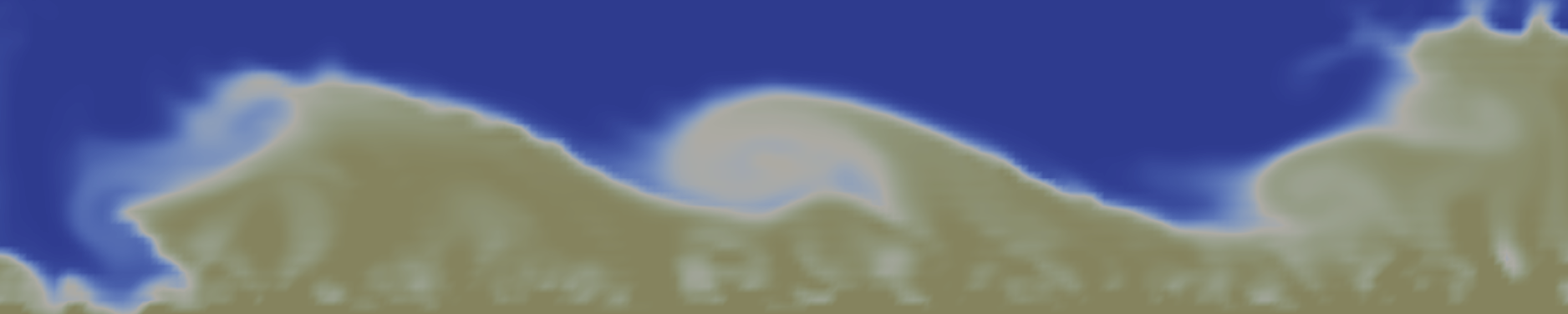}

\includegraphics[width=\columnwidth]{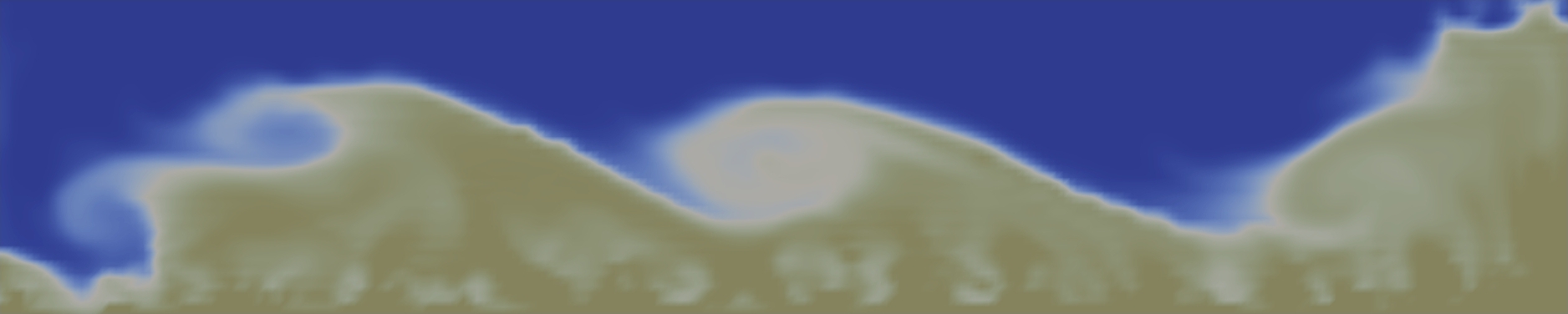}

\includegraphics[width=\columnwidth]{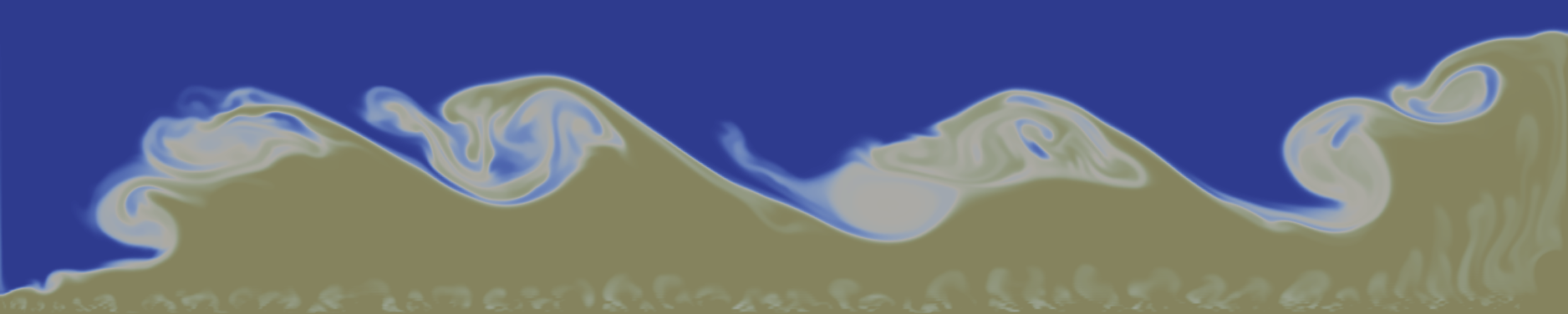}

\includegraphics[width=\columnwidth]{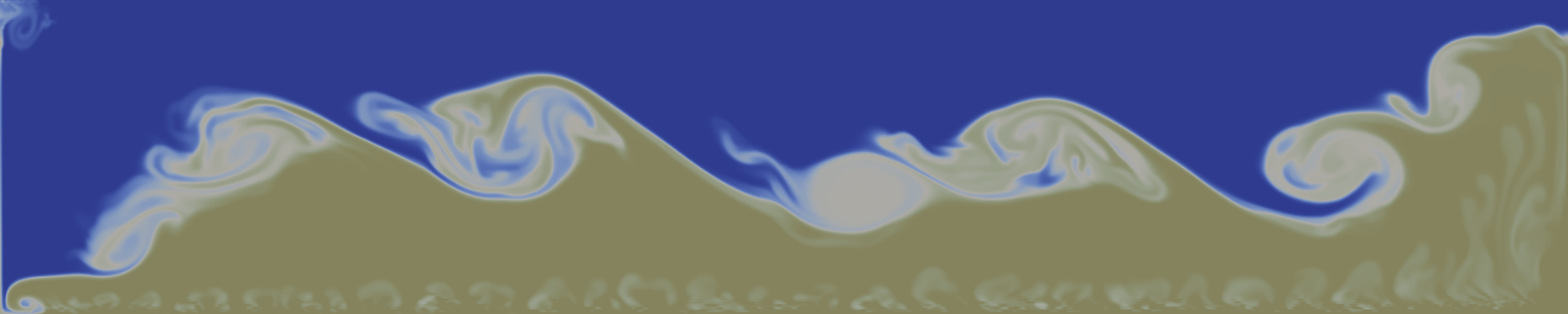}

\includegraphics[width=\columnwidth]{rhoStime1114}
\caption{Snapshots of the solid-phase bulk density at $t/T_{b}=1.114$
  evaluated with different LES models: (a) low-res Smagorinsky, (b)
  low-res one eq. eddy, (c) mid-res Smagorinsky, (d) mid-res one
  eq. eddy, (e) ultra-res DNS.}
  \label{fig:LESmodel}
\end{figure}

\begin{figure}[h]
  \centering
  \includegraphics[width=0.9\columnwidth]{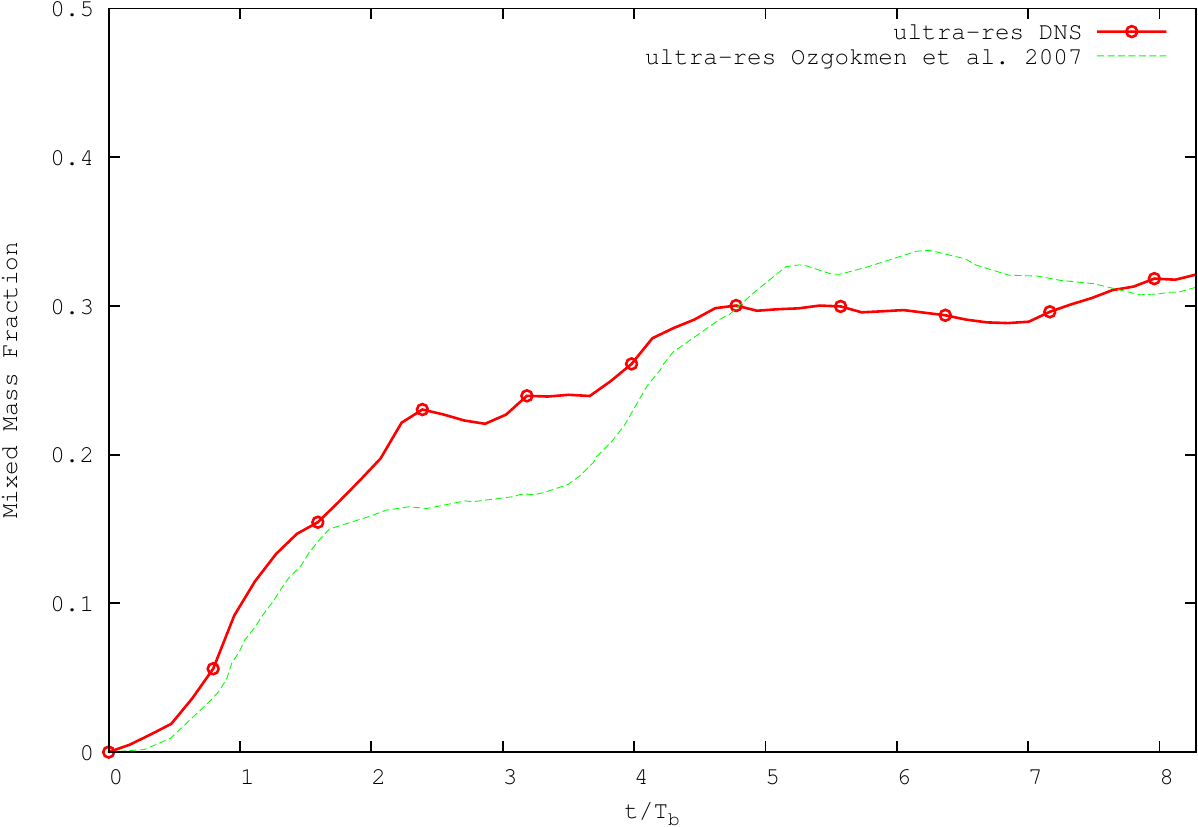}
  \caption{Time evolution of the mixed mass fraction. DNS results from
    the current study (solid) and from Ref~\cite{OIFSD2007}
    (dashed). Both simulations use the same mesh resolution.}
  \label{fig:mass-fraction}
\end{figure}

\begin{figure}[h]
  \centering
  \includegraphics[width=0.9\columnwidth]{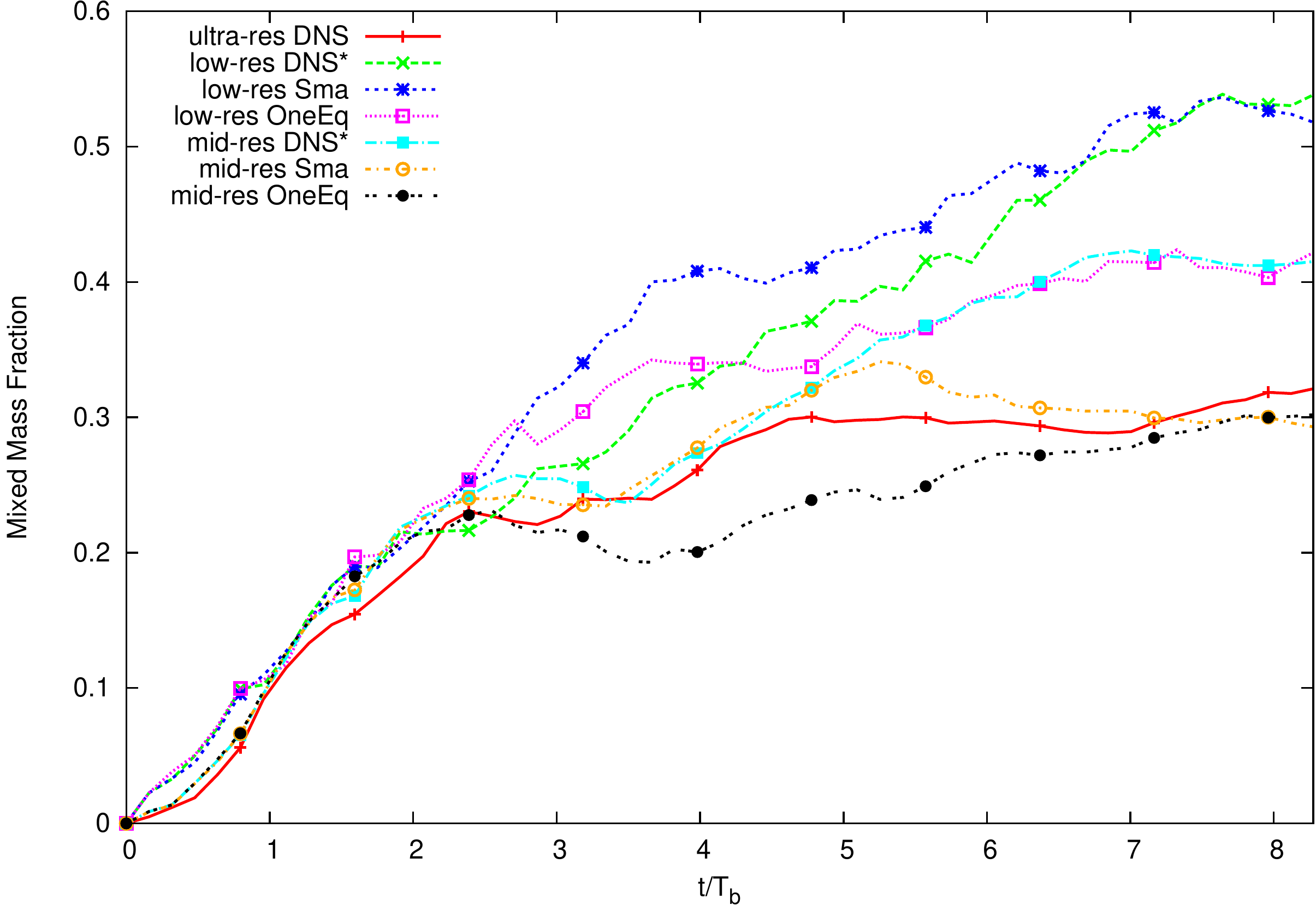}
  \caption{Time evolution of the mixed mass fraction with
    $\frac{1}{3}<\rho_{s}<\frac{2}{3}$ for the various low resolution
    LES models. The DNS results (solid) serve as benchmark.}
\label{fig:mixed-mass}
\end{figure}

\begin{figure}[h]
  \centering
\includegraphics[width=0.9\columnwidth]{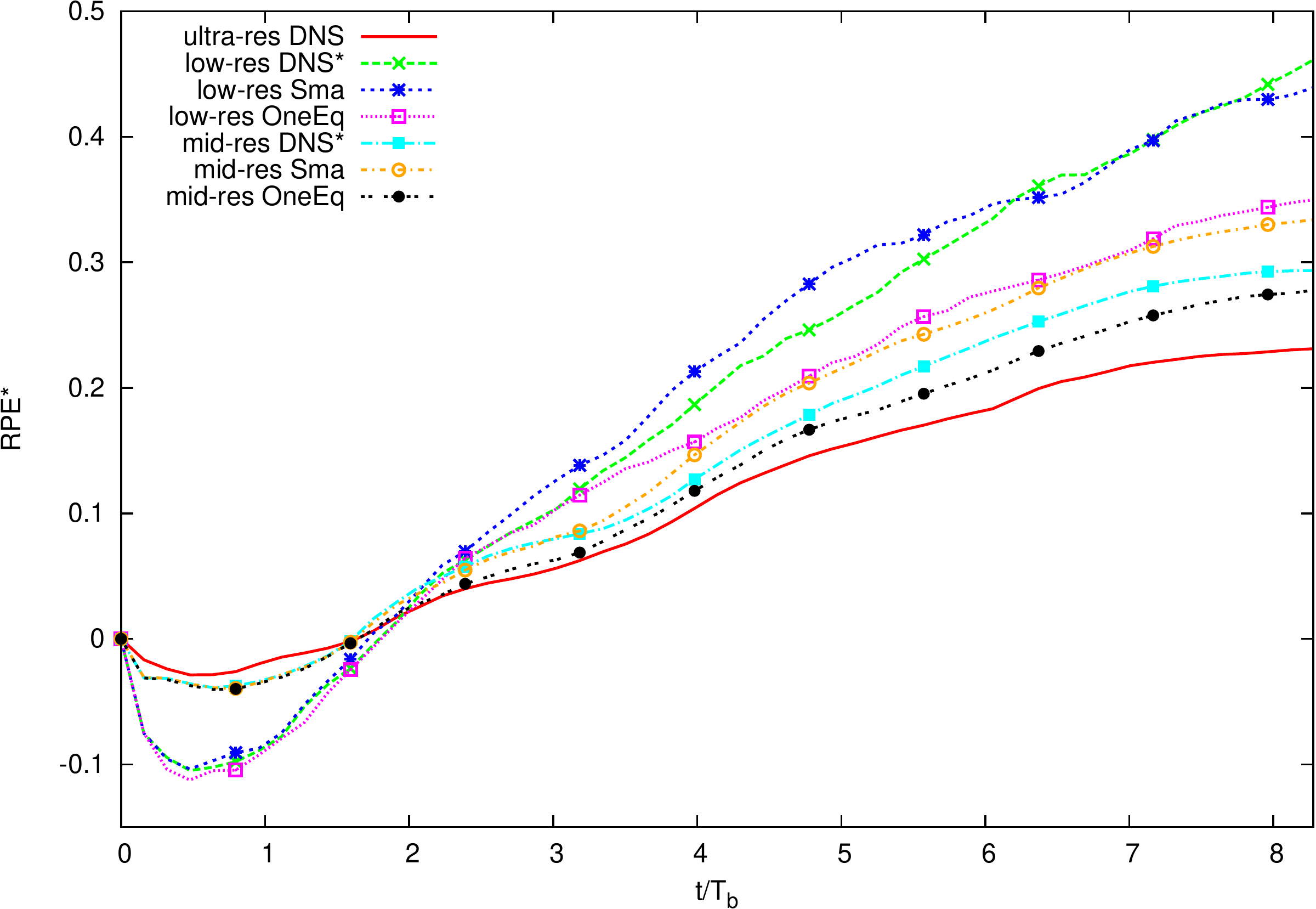}
\caption{Time evolution of the non-dimensional background energy
  ($\textup{RPE}^*$), for the LES models at various resolutions. The
  DNS results (solid) serve as benchmark. The time is normalized with
  $T_b$.}
\label{fig:BGE}
\end{figure}

The high-resolution DNS, denoted ultra-res in the remainder of the
paper, were performed on a HPC architecture (BLUGENE/Q system
installed at CINECA) with 1024 cores. These ultra-res runs took about
5~days. The medium-resolution simulations, denoted by mid-res, were
performed on 62~cores (using the HPC infrastructure of INGV, Pisa
section) for about 2~days.  Since many options for LES of compressible
multiphase flows are available, we chose to compare the ones that
OpenFOAM has built-in, to detect the most promising for our test
case. In Fig.~\ref{fig:LESmodel}-\ref{fig:mixed-mass}-\ref{fig:BGE} we
especially address this topic.  More specifically, the LES runs were
performed using either the compressible Smagorinsky model or the one
equation eddy model, that is in Eq.~\eqref{eq:equilibriumEulerian-iso}
the stress tensor $\mathbbm{T}$ is replaced by
\begin{equation*}
 \mathbbm{T}_\textup{LES} :=
  2(\mu(T) + \mu_\textup{SGS})\,\left[\frac{\nabla{u}+\nabla{u}^{T}}{2}-
  \frac{1}{D}(\nabla\cdot{u})\,\mathbbm{I}\right].
\end{equation*}
In both cases we define a subgrid-scale (SGS) stress tensor as
in~\cite{Fur1996} by
\begin{equation*}
  \mathbbm{B} = \frac{2}{D}k\,\mathbbm{I} - 2 C_k \sqrt{k}\, \delta
  \operatorname{dev}(\mathbbm{D})\,,
\end{equation*}
where $k$ is the SGS kinetic energy, $C_k = 0.02$, $\delta$ is the
grid-scale, $\mathbbm{D}= \operatorname{sym}(\nabla u$), and
$\operatorname{dev}(\mathbbm{D}) = \mathbbm{D} -
\operatorname{Tr}(\mathbbm{D})\mathbbm{I}/D$. In the Smagorinsky
model, $k$ is obtained by using the equilibrium assumption 
\begin{equation*}
\rho\,
\mathbbm{D}:\mathbbm{B} + \frac{C_e\rho}{\delta} k^{3/2} = 0\,,
\end{equation*}
where $C_e=1.048$.  Finally, the SGS viscosity is
$\mu_\textup{SGS}=C_k \rho\, \delta\, \sqrt{k}$.

On the other hand, in the one equation eddy viscosity model (which is
the compressible counterpart of the so called TKE
model~\cite{CRR2014}), $k$ is obtained through the following balance
law:
\begin{equation*}
  \partial_t(\rho\, k) + \nabla\cdot (\rho\, u\, k) - \nabla\cdot \left((\mu
    + \mu_\textup{SGS}) \nabla 
    k\right) = - \left(\rho\, \mathbbm{D}:\mathbbm{B} + \frac{C_e\rho}{\delta}
    k^{3/2}\right)\,,
\end{equation*}
keeping $\mu_\textup{SGS} = C_k \rho\, \delta\, \sqrt{k}$.  We perform
our simulations at three different resolutions, see
Table~\ref{tab:resolutions}.
\begin{table}[h!]
  \centering
\begin{tabular}{|c|l|}
  \hline
  low-res& N=10,580 
  \\
  mid-res& N=264,500
  \\
  ultra-res&N=1,058,000
  \\
  \hline
\end{tabular}
\caption{$N$ is the number of nodes of the different homogeneous meshes for our
  simulations.} 
\label{tab:resolutions}
\end{table}

Together with the DNS simulation done on the ultra-res mesh and the
four LES done on low-res and mid-res meshes, we also performed two
under-resolved simulations without SGS model, denoted by low-res DNS*
and mid-res DNS*.

\begin{figure}
\centering
\includegraphics[width=\columnwidth]{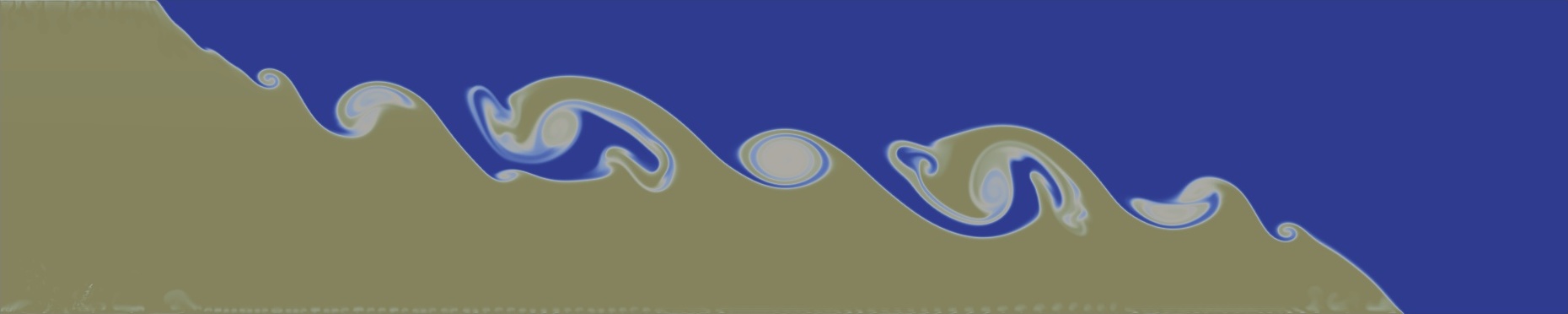}

\includegraphics[width=\columnwidth]{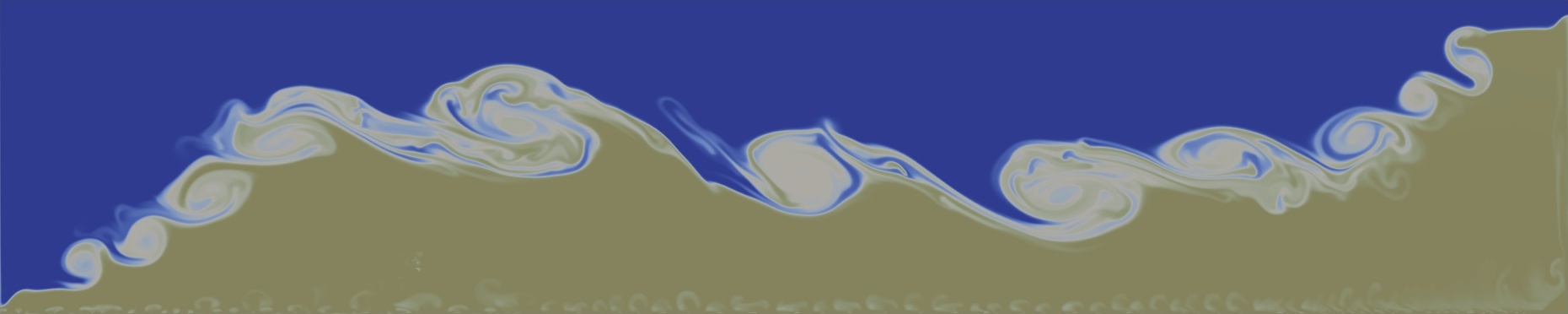}

\includegraphics[width=\columnwidth]{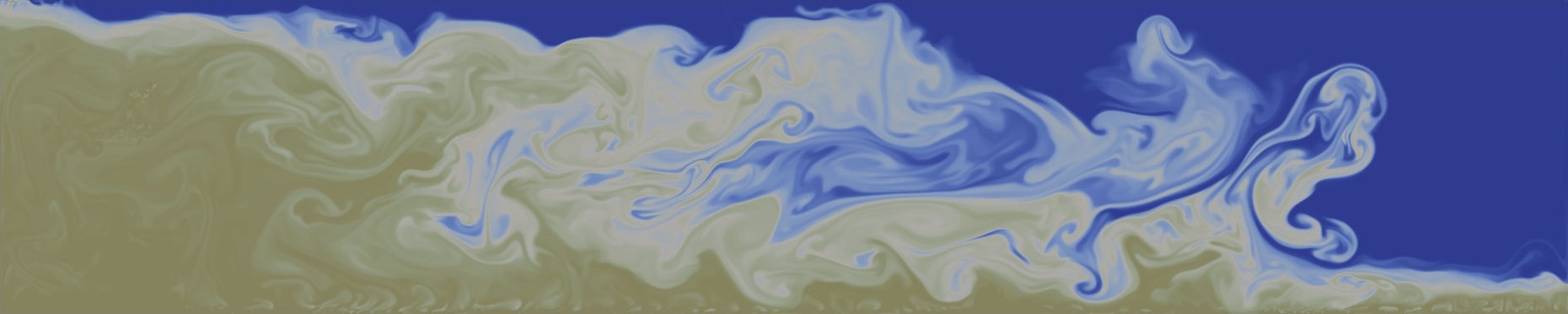}

\includegraphics[width=\columnwidth]{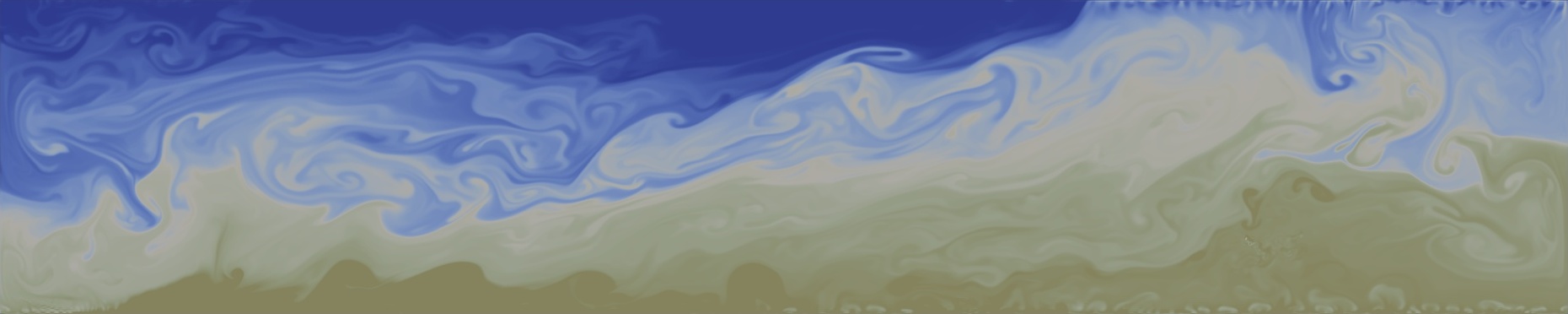}
\caption{Snapshots of the solid-phase bulk density at a) $t/T_{b}=0.637$,
b) $t/T_{b}=1.114$, c) $t/T_{b}=4.297$, d) $t/T_{b}=8.276$, in
ultra-res DNS at $Re=4300$.}
\label{fig:one}
\end{figure}

\begin{figure}
\centering
\includegraphics[width=\columnwidth]{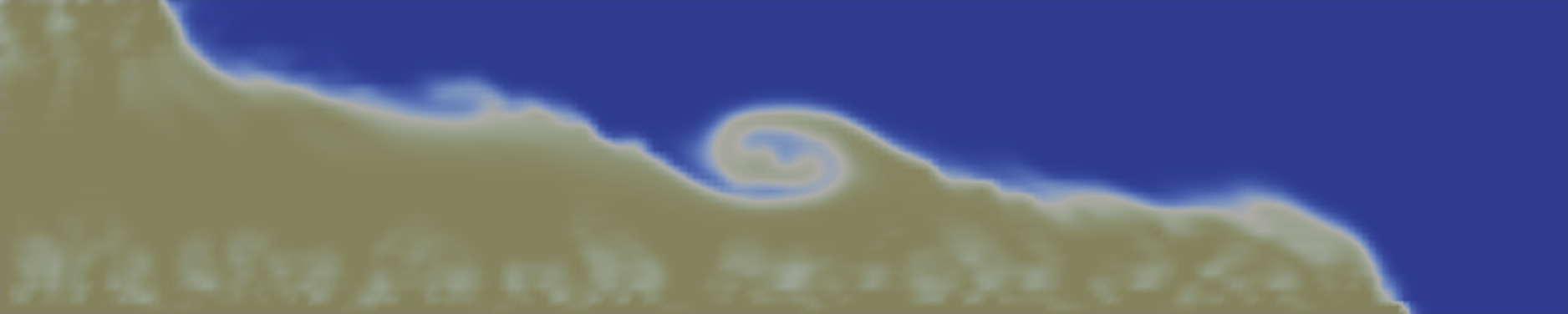}

\includegraphics[width=\columnwidth]{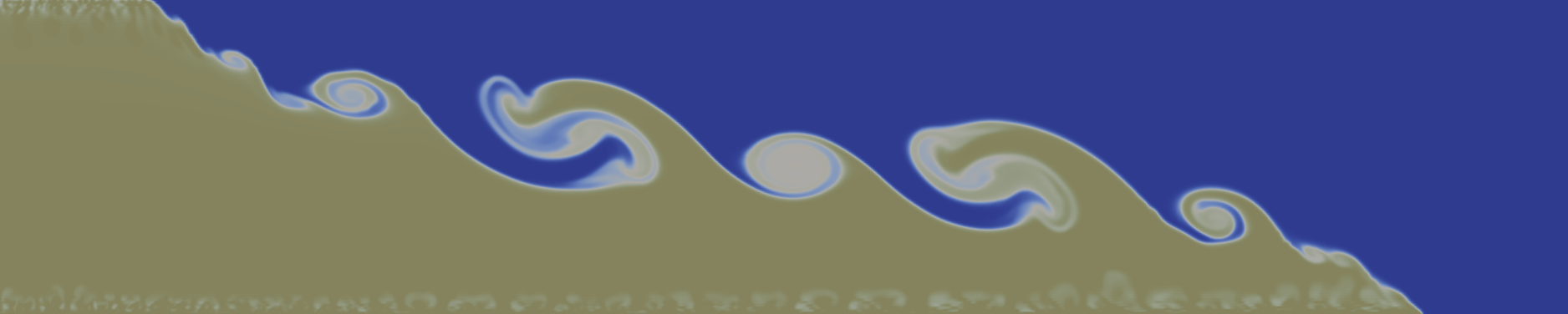}

\includegraphics[width=\columnwidth]{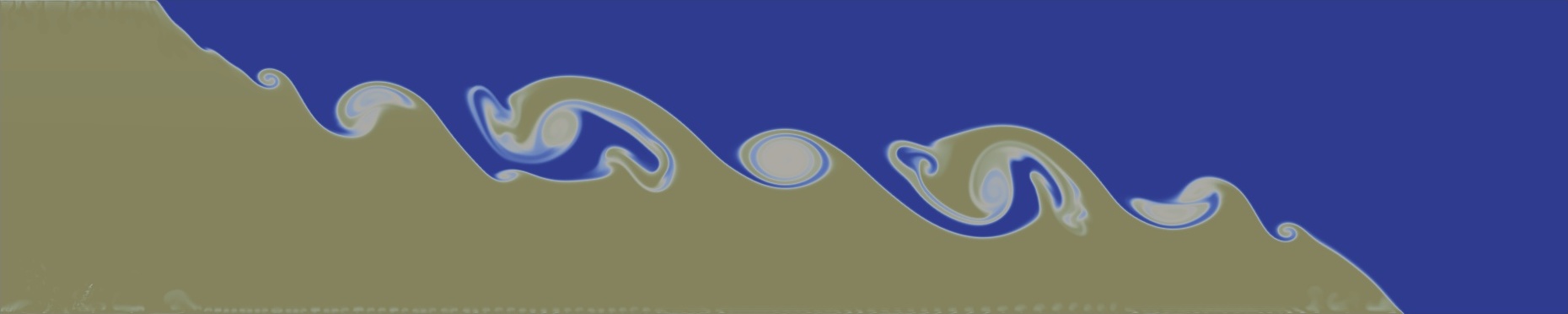}
\caption{Snapshots of the solid-phase bulk density at $t/T_{b}=0.637$
  evaluated with different resolutions, (a) low-res DNS*, (b) mid-res
  DNS*, (c) ultra-res DNS.}
  \label{fig:mesh}
\end{figure}

To illustrate the complexity of the mixing process that we
investigate, in Fig.~\ref{fig:one} we present snapshots of DNS for the
density $\rho_s$ of particles' concentration at different times (it is
represented in a linear color scale for $0\leq \rho_s \leq 1$). We
notice that the results are similar to those obtained
in~\cite{BFIO2011,OIF2009}. Thus, the DNS time evolution of the
density perturbation will be used as benchmark for other numerical
simulations, since (as in~\cite{OIFSD2007}) the number of grid points
is large enough to resolve all the relevant scales and to consider
simulations at ultra-res as a DNS.
We study this problem varying both the mesh resolution
(cf.~Table~\ref{tab:resolutions}) and the SGS LES model (Smagorinsky
and one equation eddy model). Fig.~\ref{fig:mesh} displays snapshots
of the solid-phase bulk densities at time $t=4$ for the three
different mesh resolutions: DNS at ultra-res, DNS* at mid-res, and DNS*
at low-res. Fig.~\ref{fig:LESmodel} displays snapshots of the
solid-phase bulk density at time $t = 7$. To generate the plots in
Fig.~\ref{fig:LESmodel}, we use two LES models (the Smagorinsky and
the one equation eddy model) at two coarse resolutions (mid-res and
low-res). To assess the quality of the LES results, we used the DNS at
ultra-res as benchmark.  Fig.~\ref{fig:LESmodel} shows that the LES
models yield similar results. From Fig.~\ref{fig:LESmodel} we can
deduce that, even if the overall qualitative behavior is reproduced in
four LES simulations, the results obtained at low-res are rather poor
and only the bigger vortices are reproduced. On the other hand, the
LES results at mid-res are in good agreement with the DNS and the
one equation eddy model seems to be better performing when looking at
the smaller vortices. The two LES models required a comparable
computational time and a comparison based on more quantitative
arguments will be discussed later on, see Fig.~\ref{fig:mixed-mass}
and~\ref{fig:BGE} and discussion therein.

Figs.~\ref{fig:one}-\ref{fig:LESmodel} show that, just as in the case
of the Boussinesq equations, the system rapidly generates the
Kelvin-Helmholtz billows along the interface of gravity waves, which
are counter-propagating. These waves are reflected by the side walls
and gradually both billows grow by entraining the surrounding
fluid. Later the mixing increases so much that individual billows
cannot be seen anymore.

In order to check whether our DNS results are an appropriate benchmark
for the LES results, we compare our ultra-res DNS results with those
in~\cite{OIFSD2007}. Since we chose analogous initial conditions and
since our two-phase model is comparable with the Boussinesq equations
(cf. Eq.~\eqref{eq:Boussinesq}), we expect similar qualitative results
for all the flow variables. In Fig.~\ref{fig:mass-fraction} we compare
our ultra-res DNS results with those from \cite{OIFSD2007}
using the mixed mass fraction, which is a quantity measuring the
mixing. The mixed mass fraction is defined as the fraction of volume
were the density perturbation is partially mixed. In particular, in
our simulations with homogeneous meshes, it is obtained evaluating the
percentage of cells such that $1/3<\rho_s<2/3$
(cf.~\cite{OIFSD2007}). The plots in Fig.~\ref{fig:mass-fraction} show
that the two simulations yield similar results, as expected. The main
difference is in the time interval $2<t/T_b<4$, where our simulation
seems to mix slightly more than the simulation
from~\cite{OIFSD2007}. As we will discuss later, this is probably due
to the mixing induced by the creation of stratification.
In Fig.~\ref{fig:mixed-mass} we plot the evolution of the mixed mass
fraction for all our simulations. Fig.~\ref{fig:mixed-mass} yields the
following conclusions: At the low-res, the one equation eddy model
performs the best, followed by the DNS*, and the Smagorinsky model (in
this order). At the mid-res, the Smagorinsky model performs the best,
followed by the one equation eddy model, and the DNS* (in this order).

The main measure used in the assessment of the accuracy of the models
employed to predict mixing in the dam-break problem is the
non-dimensional background potential energy RPE* defined
in~\eqref{rpenondimensional}, cf.~\cite{WLRD1995}.
Figure~\ref{fig:BGE} plots the background energy of the various LES
models. The DNS results serve as benchmark.  Fig.~\ref{fig:BGE} yields
the following conclusions: At the low-res, the one equation eddy model
performs the best, followed by the DNS*, and the Smagorinsky model (in
this order).  At the mid-res, the one equation eddy model again
performs the best, followed by the DNS*, and the Smagorinsky model (in
this order).

Apart from the above LES model assessment, we also observe that new
important phenomena appear in the compressible case: While in the
incompressible case the RPE is monotonically increasing, in our
investigation it is initially decreasing, it then reaches a minimum,
and it finally starts to increase monotonically, as expected. In order
to better understand this phenomenon, we have to compare the
background energy of the homogeneous initial condition with that of
the stratified initial condition. Evaluating the initial potential
energy ($\textup{PE}_0$), the available energy ($\textup{APE}_0$), and
the background energy ($\textup{RPE}_0$) for the homogeneous initial
density of the solid-phase, we get:
\begin{align}
  \label{eq:iniHom}
  \textup{PE}_0/g & = 10, & \textup{APE}_0 & = 5, & \textup{RPE}_0/g =
  5.
\end{align}
If we consider the initial distribution of fluid and particles in the
stratified case, with $\rho_s(x,z,0)=0.01\,\rho_f(x,z,0)$, and
$\rho_{f,\textup{homog.}} = 100$, we get
\begin{equation}
  \label{eq:rhoSstrat} 
  \rho_{s}(x,z)=\frac{1-\eta\,
    z}{1-\eta \frac{H}{2}} \mathcal{H}(-x)\,,
\end{equation}
where $\mathcal{H}(x)$ is the Heaviside step function. Evaluating the
same energies (as those in~\eqref{eq:iniHom}) for the stratified
density distribution considered and using $H=2$ and $L=10$, we get
\begin{align}
  \textup{PE}_\textup{str.}/g & = \frac{10}{3}\frac{(3 - 4\eta)}{1 -
    \eta}, & \textup{APE}_\textup{str.} & = \frac{1}{2}
  \textup{PE}_\textup{str.}, & \textup{RPE}_\textup{str.} =
  \frac{1}{2} \textup{PE}_\textup{str.}\,,
\end{align}
and also
\begin{equation}
  \textup{RPE}_\textup{str.}^* = \frac{\textup{RPE}_\textup{str.} -
    \textup{RPE}_0}{\textup{RPE}_0} = 
  \frac{-\eta}{3(1-\eta)} < 0\,.
\end{equation}
These analytical computations show that the RPE of the stratified
state is smaller than that of the homogeneous state. In the next
section we will discuss this issue in more detail.
\subsection{A few remarks on the model without the barotropic
  assumption.}
In this section, we compare the results of the previous sections with
some low-res simulations obtained from the same test case, by using
system~\eqref{eq:equilibriumEulerian2bis}, i.e. without the assumption
of a barotropic fluid. The simulations with
model~\eqref{eq:equilibriumEulerian2bis} are more time-consuming and
so we performed them only at low-res (simulations with finer mesh
resolution are in preparation and their results will appear in the
forthcoming report~\cite{Cer2014}).

\begin{figure}[h]
\centering
\includegraphics[width=0.9\columnwidth]{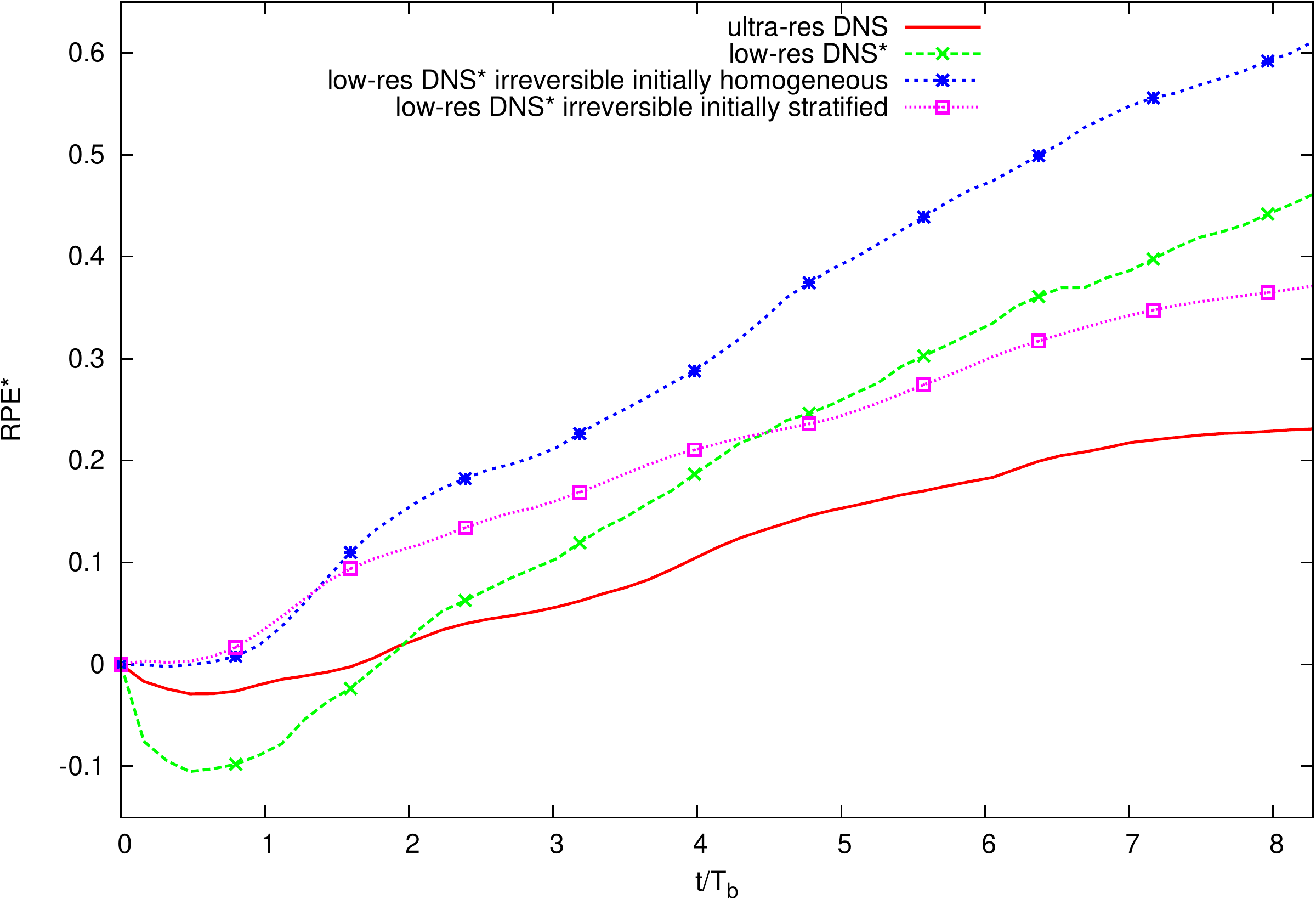}
\caption{Plot of RPE* obtained by using
  model~\eqref{eq:equilibriumEulerian2bis}. Low res DNS* compared with
  the ultra-res DNS and the low-res DNS*. The line with
  ``$\cdot\!\cdot\!\square\!\cdot\!\cdot$'' represents RPE* starting
  from the initial condition~\eqref{eq:rhoSstrat}, while the line with
  ``- -$\ast$- -'' represents the same quantity starting from the
  homogeneous initial state. The solid line and the line with ``- -$\times$-
  -'' are the RPE* obtained with the barotropic
  model~\eqref{eq:equilibriumEulerian-iso} with homogeneous initial
  state, with the ultra-res DNS and the low-res DNS*, respectively.}
\label{fig:BGE_irreversible}  
\end{figure}

The barotropic assumption is based on the fact that the thermal and
kinematic diffusion ($\nabla \cdot q$ and $\mathbb{T}:\nabla u$) in
Eq.~\eqref{eq:equilibriumEulerian2bis} are negligible, so that the
entropy $s$ of the system is constant along streamlines, i.e.
\mbox{$(\partial_t + u\cdot\nabla) s(x,z,t) = 0$} (cf.~\cite{Fei2004}
for the one-phase case and~\cite{Cer2014} for the multiphase case):
This is a reversibility assumption.  Indeed, the background energy can
be considered as a sort of entropy, measuring the potential energy
dispersed in the mixing~\cite{WLRD1995}. The fact that the
transformation is reversible allows the background energy to
decrease. On the contrary, if we remove this assumption, coming back
to the full multiphase model~\eqref{eq:equilibriumEulerian2bis}
(including the energy equation), we find that the background energy
becomes monotone, see Fig.~\ref{fig:BGE_irreversible}. This figure
suggests that the barotropic assumption may be not completely
justified during the initial time-interval needed to adjust from the
homogeneous to the stratified condition (probably this transformation
can not be considered fully iso-entropic). Nevertheless, the
barotropic assumption seems justified after the time $T_a$.

Moreover, the stratified initial condition makes the simulation more
stable and accurate, but also less diffusive, even at low-res. The
RPE* is monotonically increasing when using
model~\eqref{eq:equilibriumEulerian2bis} (low-res irreversible) and,
starting with the stratified initial condition, decreases the mixing
and brings it closer to that of the DNS.

Note that the low-res DNS* irreversible with homogeneous initial data
and the ultra-res DNS start from the same datum. Even if the low-res
DNS* is under-resolved, the behavior of the RPE* is correct and it is
monotonically increasing. The behavior, at the beginning of the
evolution, is closer to the DNS than the behavior of the LES described
in Fig.~\ref{fig:BGE}, obtained from the barotropic
model~\eqref{eq:equilibriumEulerian-iso}.  On the other hand, after
this transient time the behavior becomes comparable with that of the
previous low-res barotropic simulation (low-res DNS* vs. low-res DNS*
irreversible and homogeneous).  The comparison of the results obtained
at various resolutions and with different LES models for the
barotropic and non-barotropic equations deserves further
investigation and we plan to perform it in the near future.
\section{Conclusions}
We examined a two-dimensional dam-break problem were the instability
is due to the presence of a dilute suspension of particles in half of
the domain. The Reynolds number based on the typical gravity wave
velocity and on the semi-height of the domain is $4300$, the Froude
number is $2^{-\frac{1}{2}}$, the Mach number is $10^{-2}$, and the
Prandtl number is $1$. The particle concentration is $10^{-3}$, and
the Stokes number is smaller than $10^{-3}$ (fine particles).  The
importance of stratification, measured as the density gradient times
the domain height ($-\partial_y\rho_f/\rho_f\,H$), is about a few
percent ($\sim5\%$). Even if the problem is quasi-incompressible and
quasi-isothermal, we used a full compressible code, with a barotropic
constitutive law. We employed a homogeneous and orthogonal mesh with
three different grid refinements ranging from $10^4$ to $10^6$
cells. \textit{A posteriori} tests confirm that the finer grid can
resolve all the scales of the problem.  The code that we used was
derived from the OpenFOAM$^{\textup{\textregistered}}$ \verb!C++!
libraries.

We compared our quasi-isothermal two-phase simulations with the
analogous mono-phase problem, where the mixing occurs between the same
fluid at two different temperatures, as reported
in~\cite{OIFSD2007}. As we showed in Section~\ref{sec:models}, this is
possible since the two physical problems become mathematically
equivalent in the regimes under study. As expected, we found a good
agreement between the two sets of numerical results. We reported the
evolution of the background (or reference) potential energy (RPE), a
scalar quantity measuring the mixing between the two fluids. The main
contributions of this report are the following: We implemented a
multiphase Eulerian model (that can be used in more complex physical
situations, with more than two phases, and also involving chemical
reactions between species, as in volcanic eruptions). We also showed
the effectiveness of the numerical results obtained programming with
an open-source code. More importantly, we discovered that peculiar
effects due to compressibility influence the mixing.  In the
literature we found that the mono-phase, incompressible Boussinesq
test case has a monotonically increasing RPE. On the other hand, in
our numerical experiments with slightly compressible two-phase flow,
we found that the RPE initially decreases because of the
stratification instability, and then it increases monotonically
because of the mixing between the particles and the surrounding
fluid. Indeed, even if the flow is quasi-incompressible ($\textup{Ma}=0.01$),
it turns out that stratification effects are not negligible.  We
reported the preliminary results in the two-dimensional case. We plan
to perform three-dimensional numerical simulations of the same problem
in a future study.
\def\ocirc#1{\ifmmode\setbox0=\hbox{$#1$}\dimen0=\ht0 \advance\dimen0
  by1pt\rlap{\hbox to\wd0{\hss\raise\dimen0
  \hbox{\hskip.2em$\scriptscriptstyle\circ$}\hss}}#1\else {\accent"17 #1}\fi}
  \def\polhk#1{\setbox0=\hbox{#1}{\ooalign{\hidewidth
  \lower1.5ex\hbox{`}\hidewidth\crcr\unhbox0}}} \def\cprime{$'$}


\begin{thebibliography}{10}
%
\bibitem{BE2010}
S.~Balachandar and J.K. Eaton.
\newblock Turbulent dispersed multi-phase flow.
\newblock \textit{Annu.
  Rev. Fluid Mech.}, vol.~42, pp.~399--434. Annual Reviews, Palo Alto, CA, 2010.


\bibitem{BFIO2011}
L.C. Berselli, P.~Fischer, T.~Iliescu, and T.~\"{O}zg\"{o}kmen.
\newblock Horizontal {L}arge {E}ddy {S}imulation of stratified mixing in a
  lock-exchange system.
\newblock \textit{J. Sci. Comput.}, 49:3--20, 2011.

\bibitem{BS1978}
R.~E. Britter and J.E. Simpson.
\newblock Experiments on the dynamics of a gravity current head.
\newblock \textit{J. Fluid Mech.}, 88:223--240, 1978.

\bibitem{Car1958}
G.F. Carrier.
\newblock Shock waves in a dusty gas.
\newblock
 \textit{J. Fluid Mech}, 4:376--382, 1958.

\bibitem{Cer2014}
M.~Cerminara.
\newblock \textit{Multiphase flows in volcanology}.
\newblock PhD thesis, Scuola Normale Superiore, 2014.
\newblock To appear.

\bibitem{CBEOS2013}
M.~Cerminara, L.C.~Berselli, T.~Esposti~Ongaro, and M.V.~Salvetti.
\newblock Direct numerical simulation of a compressible multiphase flow through
  the eulerian approach.
\newblock In \textit{Direct and Large-Eddy Simulation IX,} vol.~12 of
  \textit{ERCOFTAC Series}. Springer, 2013.
\newblock At press.

\bibitem{CRR2014}
T.~Chac\'{o}n Rebollo and R.~Lewandowski.
\newblock \textit{Mathematical and numerical foundations of turbulence
  models and applications}.
\newblock Birkh\"{a}user, Boston, 2014.

\bibitem{CRB2011}
B.~Cushman-Roisin and J.-M.~Beckers.
\newblock \textit{Introduction to Geophysical Fluid Dynamics}.
\newblock Academic Press, 2nd edition,
  2011.
\newblock ISBN: 978-0-12-088759-0.

\bibitem{Dor1998}
\"{A}.~D\"{o}rnbrack.
\newblock Turbulent mixing by breaking gravity waves.
\newblock \textit{J. Fluid Mech.} 375:113--141, 1998.

\bibitem{EOCENS2007}
T.~Esposti~Ongaro, C.~Cavazzoni, G.~Erbacci, A.~Neri, and M.V.~Salvetti.
\newblock A parallel multiphase flow code for the 3d simulation of explosive
  volcanic eruptions.
\newblock \textit{Parallel Comput.}, 33(7-8):541--560, 2007.

\bibitem{Fei2004}
E.~Feireisl.  
\emph{Dynamics of viscous compressible fluids}, Oxford University Press,
  Oxford, 2004.

\bibitem{Fer2000}
H.J.S.~Fernando.
\newblock Aspects of stratified turbulence.
\newblock In: Kerr, R.M., Kimura, Y. (Eds.),
Developments in Geophysical Turbulence, pp.~81--92, 2000.

\bibitem{FP1999}
J.H.~Ferziger and M.~Peri{\'c}. \emph{Computational methods for fluid
  dynamics}, revised ed., Springer-Verlag, Berlin, 1999.

\bibitem{Fri1995}
U.~Frisch.
\newblock \textit{Turbulence, The {L}egacy of {A}.{N}.~{K}olmogorov}.
\newblock Cambridge University Press, Cambridge, 1995.

\bibitem{Fur1996}
C.~Fureby.
\newblock On subgrid scale modeling in large eddy simulations of compressible
  fluid flow.
\newblock \textit{Phys. Fluids}, 8(5):1301--1311, 1996.

\bibitem{HLD1996}
J.~Hacker, P.~F. Linden, and S.~B. Dalziel.
\newblock Mixing in lock-release gravity currents.
\newblock \textit{Dyn. Atmos. Oceans}, 24(1-4):183--195, 1996.

\bibitem{HHPS1996}
M.A.~Hallworth, H.E.~Huppert, J.C.~Phillips, and R.S.J.~Sparks.
\newblock Entrainment into two-dimensional and axisymmetric turbulent gravity
  currents.
\newblock \textit{J. Fluid Mech.}, 308:289--311, 1996.

\bibitem{HPHS1993}
M.A.~Hallworth, J.C.~Phillips, H.E.~Huppert, and R.S.J.~Sparks.
\newblock Entrainment in turbulent gravity currents.
\newblock \textit{Nature}, 362:829 -- 831, 1993.

\bibitem{Iss1986}
R.I.~Issa.
\newblock Solution of the implicitly discretised fluid flow equations
by operator-splitting.
\newblock \textit{J. Comput. Phys.}, {62}(1):40--65, 1986.

\bibitem{Jas1996}
H.~Jasak.
\newblock \textit{Error Analysis and Estimation for the Finite Volume
  Method with Applications to Fluid Flows}.
\newblock PhD thesis, Imperial College, London, 1996.

\bibitem{KC2000}
L.H.~Kantha and C.A.~Clayson.
\newblock \textit{Small Scale Processes in Geophysical Fluid Flows}, vol.~67 of
  \textit{Int. Geophysics Series}.
\newblock Academic Press, 2000.

\bibitem{KGGS}
D.A.~Kay, P.M.~Gresho, D.F.~Griffiths, and D.J.~Silvester.
\newblock Adaptive time-stepping for incompressible flow. {II}.
  {N}avier-{S}tokes equations.
\newblock \textit{SIAM J. Sci. Comput.}, 32(1):111--128, 2010.

\bibitem{Mar1970}
F.~Marble.
\newblock Dynamics of dusty gases.
\newblock  \textit{Annu.
  Rev. Fluid Mech.}, vol.~3, pp.~397--446. Annual Reviews, Palo Alto, CA,
1970.

\bibitem{M1956}
  B. R.~Morton, G.~Taylor, and J.S.~Turner.
  \newblock Turbulent gravitational
    convection from maintained and instantaneous sources.
    \newblock    \textit{Proc. R. Soc. Lond. A}, {234}, 1--23 1956.
    
 \bibitem{OIF2009b}
T.~\"{O}zg\"{o}kmen, T.~Iliescu, and P.~Fischer.
\newblock Large eddy simulation of stratified mixing in a three-dimensional
    lock-exchange system.
\newblock \textit{Ocean Modelling}, 26:134--155, 2009.

\bibitem{OIF2009}
T.~{\"O}zg{\"o}kmen, T.~Iliescu, and P.~Fischer.
\newblock Reynolds number dependence of mixing in a lock-exchange system from
  direct numerical and large eddy simulations.
\newblock \textit{Ocean Modelling}, 30(2):190--206, 2009.

\bibitem{OIFSD2007}
T.~\"{O}zg\"{o}kmen, T.~Iliescu, P.~Fischer, A.~Srinivasan, and J.~Duan.
\newblock Large eddy simulation of stratified mixing in two-dimensional
  dam-break problem in a rectangular enclosed domain.
\newblock \textit{Ocean Modelling}, 16:106--140, 2007.

\bibitem{RL2000}
J.J.~Riley and M.-P.~Lelong.
\newblock Fluid motions in presence of strong stable stratification.
\newblock In \textit{Annu. Rev. Fluid Mech.}, vol.~32, pp.~613--657.
Annual Reviews, Palo Alto, CA, 2000.

\bibitem{SD1994}
D.A.~Siegel and  J.A.~Domaradzki.
\newblock Large-eddy  simulation of  decaying  stably stratified turbulence.
\newblock \textit{J. Phys. Oceanogr.}, 24:2353--2386, 1994.


\bibitem{SKOH2005}
Y.J.~Suzuki, T.~Koyaguchi, M.~Ogawa, and I.~Hachisu.
\newblock A numerical study of turbulent mixing in eruption clouds using a
  3D fluid dynamics model.
\newblock \textit{J. Geophys. Res.: Solid Earth}, 110(B8):B08201,
  2005.
  
\bibitem{VHC2014}
 S.A.~Valade, A.J.L.~Harris and M.~Cerminara.
\newblock Plume Ascent Tracker: Interactive Matlab software for
analysis of ascending plumes in image data.
\newblock \textit{Comput. \& Geosci.}, 66(0):132--144, 2014.

\bibitem{WLRD1995}
K.B.~Winters, P.N.~Lombard, J.J.~Riley, and E.A.~D'Asaro.
\newblock Available potential energy and mixing in density-stratified fluids.
\newblock \textit{J. Fluid Mech.}, 289:115--128, 4 1995.
%
\end{thebibliography}
\end{document}